\documentclass[12pt]{iopart}
\usepackage{graphicx}

\begin{document}
\title[Optimized fabrication of high quality La$_{0.67}$Sr$_{0.33}$MnO$_3$ thin films]{Optimized fabrication of high quality La$_{0.67}$Sr$_{0.33}$MnO$_3$ thin films considering all essential characteristics}
\author{H Boschker$^1$, M Huijben$^1$, A Vailionis$^{2,3}$, J Verbeeck$^4$, S van Aert$^4$, M Luysberg$^5$, S Bals$^4$, G van Tendeloo$^4$, E P Houwman$^1$, G Koster$^1$, D H A Blank$^1$ and G Rijnders$^1$}
\address{$^1$ Faculty of Science and Technology and MESA$^+$ Institute for Nanotechnology, University of Twente, 7500 AE, Enschede, The Netherlands}
\address{$^2$ Geballe Laboratory for Advanced Materials, Stanford University, Stanford, California 94305, USA}
\address{$^3$ Stanford Institute for Materials and Energy Sciences, SLAC National Accelerator Laboratory, 2575 Sand Hill Road, Menlo Park, California 94025, USA}
\address{$^4$ Electron Microscopy for Materials Science (EMAT), University of Antwerp, Groenenborgerlaan 171, 2020 Antwerp, Belgium}
\address{$^5$ Institute of Solid State Research and Ernst Ruska Center for Microscopy and Spectroscopy with Electrons, Helmholtz Research Center J\"{u}lich, 52425 J\"{u}lich, Germany}
\ead{a.j.h.m.rijnders@utwente.nl}
\begin{abstract}
In this article, an overview of the fabrication and properties of high quality La$_{0.67}$Sr$_{0.33}$MnO$_3$ (LSMO) thin films is given. A high quality LSMO film combines a smooth surface morphology with a large magnetization and a small residual resistivity, while avoiding precipitates and surface segregation. In literature, typically only a few of these issues are adressed. We therefore present a thorough characterization of our films, which were grown by pulsed laser deposition. The films were characterized with reflection high energy electron diffraction, atomic force microscopy, x-ray diffraction, magnetization and transport measurements, x-ray photoelectron spectroscopy and scanning transmission electron microscopy. The films have a saturation magnetization of 4.0 $\mu_\textrm{B}$/Mn, a Curie temperature of 350 K and a residual resistivity of 60 $\mu\Omega$cm. These results indicate that high quality films, combining both large magnetization and small residual resistivity, were realized. A comparison between different samples presented in literature shows that focussing on a single property is insufficient for the optimization of the deposition process. For high quality films, all properties have to be adressed. For LSMO devices, the thin film quality is crucial for the device performance. Therefore, this research is important for the application of LSMO in devices.
\end{abstract}
\submitto{\JPD}
\maketitle

\section{Introduction}
The discovery of colossal magnetoresistance in the perovskite manganites in 1993 resulted in an extensive amount of research \cite{Helmolt1993, Jin1994}, because of the potential for data storage applications. Among the perovskite manganites, La$_{0.67}$Sr$_{0.33}$MnO$_3$ (LSMO) has the largest single electron bandwidth and the highest Curie temperature. These properties, together with the 100\% spin polarization, make LSMO an interesting material for application in spintronic devices. Examples are magnetic tunnel junctions \cite{Sun1996, Odonnell2000, Bowen2003, Ogimoto2003}, Schottky devices \cite{Postma2004, Nakagawa2005, Hikita2009} and magnetoelectric devices \cite{Eerenstein2007, Molegraaf2009, Garcia2009, Yu2010, Wu2010, Borisevich2010}. Furthermore, LSMO is also used to investigate other materials, e.g. by spin injection into cuprate superconductors \cite{sawa2000, Chen2001, Pena2005, vanZalk2009}, to probe spin polarization at the BaTiO$_3$/Fe interface \cite{Garcia2010} and to study spin dependent transport in organic materials \cite{Barraud2010}. 

All these applications require LSMO thin films and the device performance will mainly depend on the thin film quality. The optimization of the LSMO thin films is not straightforward, as different parameters exist which can be optimized, e.g. the magnetization and the electrical conductivity. In addition to these parameters, a smooth surface morphology is required for most applications. Focussing on ideal layer-by-layer growth can compromise the functional properties of the film \cite{Huijben2008}. Other detrimental effects are Mn$_3$O$_4$ precipitates on LSMO surfaces, which are the result of off-stoichiometric deposition \cite{Higuchi2009}, and Sr segregation towards the surface \cite{Herger2008}.

Summarizing, a high quality LSMO film combines an atomically smooth surface morphology with a large magnetization and a low residual resistivity, while avoiding precipitates and surface segregation. In literature, typically only a few of these issues are adressed, e.g., by focussing only on the electrical conductivity. Here, we show that high quality films, combining good ferromagnetic properties with a high conductivity can be realized. As discussed in a previous paper \cite{Huijben2008}, we optimized the LSMO films on a combination of the magnetic properties and two dimensional (2D) layer-by-layer growth. We found that the properties of the films are very sensitive to the amount of oxygen present during the deposition. Too little oxygen resulted in inferior ferromagnetic properties, while too much oxygen resulted in three dimensional growth. In this paper, we present a thorough characterization of the resulting films and compare them to examples found in literature. The films have the highest reported saturation magnetization. The optimization on ferromagnetism and 2D layer-by-layer growth resulted in a high conductivity and a lack of surface segregation as well.

This article is organized as follows. First, the growth of the films is described, together with an analysis of the surface morphology with reflection high energy electron diffraction (RHEED) and scanning probe microscopy (SPM). Then the crystal structure is discussed, followed by the functional properties, magnetization and electrical conductivity. In the next section, \textit{in situ} x-ray photoemission spectroscopy (XPS) measurements are presented, which are used to study the possibility of surface segregation. The interface atomic structure was measured with scanning transmission electron microscopy (STEM). In the final section, the properties of the LSMO thin films are compared with films described in literature.

\section{Thin film growth}
\label{lsmogrowth}
In this section, the growth of the LSMO thin films is described. First, the treatment of the substrates used as template for the growth is discussed. Then, the growth of the thin films by pulsed laser deposition is described. \textit{In situ} RHEED measurements are presented from which it is concluded that the initial growth occurs in the two dimensional layer-by-layer growth mode, and that a transition to a steady state growth mode occurs for thicker films. Finally, the surface morphology was characterized with scanning probe microscopy, which showed smooth surfaces without the presence of precipitates. 

\subsection{Substrates}
Several substrates are commercially available for the growth of oxide thin films. A number of substrates which have comparable lattice 
parameters to LSMO can be used as a template for the growth of the LSMO thin films. They include NdGaO$_3$ (NGO), (LaAlO$_3$)$_{0,3}$-(Sr$_2$AlTaO$_6$)$_{0,7}$ (LSAT), SrTiO$_3$ (STO) and DyScO$_3$ (DSO). The pseudocubic lattice parameters are respectively 3.85, 3.87, 3.905 and 3.95~\AA. STO and LSAT are cubic crystals while NGO and DSO have an orthorhombic crystal structure. For the studies in this article, we used STO (001)$_\textrm{c}$ substrates, as STO is the standard substrate used for LSMO growth in literature and this allows for better comparison between the samples. (The subscripts o, c and pc denote the orthorhombic, cubic and pseudocubic crystal structure, respectively.)

The substrates were first ultrasonically cleaned with acetone and ethanol. Singly TiO$_2$ terminated ($B$-site) STO (001)$_\textrm{c}$ substrates were obtained with the procedure developed by Koster \textit{et al.} \cite{Koster1998}. In some cases, the annealing step of 2 hours at 950$^\circ$C resulted in some Sr segregation from the bulk towards the surface, due to reduced STO quality \cite{Bachelet2009, Kleibeuker2010}. This could be observed with atomic force microscopy (AFM), either in the phase contrast or with the presence of straight stepedges aligned with the principal crystal directions \cite{Koster2000}. These substrates were treated with H$_2$O and the HF solution again, followed by a 30 minute anneal at 920$^\circ$C, after which no SrO termination could be detected. 

\subsection{Pulsed laser deposition}
The LSMO thin films were grown with pulsed laser deposition (PLD) (TSST system). The substrate temperature during growth was 750-800 $^{\circ}$C in an oxygen environment of 0.27 mbar. The laser beam was produced by a 248-nm-wavelength KrF excimer laser (LPXPro$^\textrm{TM}$ from Coherent, Inc.) with a typical pulse duration of 20-30 ns. With a 4 by 15 mm rectangular mask the most homogeneous part of the laser beam was selected. An image of the mask was created on the stoichiometric target (Praxair electronics) with a lens, resulting in a spotsize of 2.3 mm$^2$ (0.9 by 2.5 mm). The beam energy was controlled with a variable attenuator or with the laser voltage, yielding a fluence at the target of 2 J/cm$^2$. The repetition rate was 1 Hz and the substrate was placed at 5 cm distance directly opposite to the target. Before deposition, the target was pre-ablated for 2 minutes at 5 Hz to remove any possible surface contamination. After deposition, the PLD chamber was flooded with pure oxygen (typically 100 mbar) and the samples were cooled down by switching of the heater power. Typically, the cooldown required 2 hours.

\subsection{Growth mode}
\begin{figure}
\centering
\includegraphics*[width=8cm]{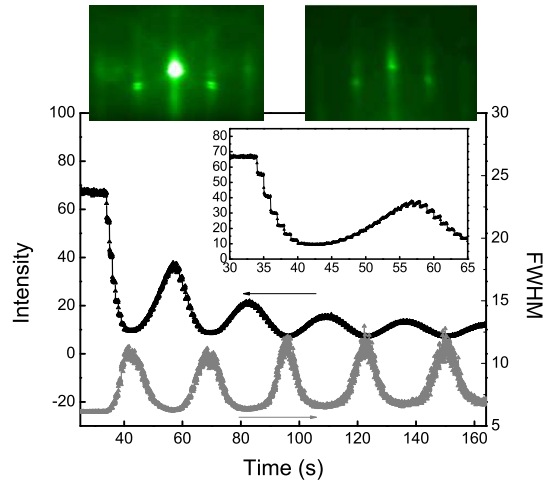}
\caption{RHEED specular spot intensity and FWHM as monitored during the initial growth of LSMO on STO (001)$_\textrm{c}$. The inset shows an expanded view of the growth of the first unit cell layer. Two RHEED images are presented, the left image was taken before deposition (at low pressure) and the right image was taken after deposition of 5 unit cell layers. }
\label{rheedini}
\end{figure}

\begin{figure}
\centering
\includegraphics*[width=8cm]{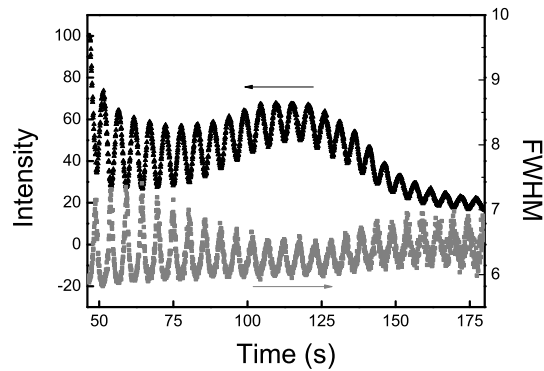}
\caption{RHEED specular spot intensity and FWHM as monitored during the growth of a relatively thick film of LSMO on STO (001)$_\textrm{c}$. The graph shows the 7$^\textrm{th}$ to 32$^\textrm{nd}$ oscillation. A recovery of the RHEED oscillation intensity maximum, followed by a rapid decrease, is observed around the 20$^\textrm{th}$ unit cell layer.}
\label{rheed25}
\end{figure}

The growth of the films was studied \textit{in situ} with RHEED during the growth. The substrate RHEED pattern is shown in figure~\ref{rheedini} on the top left. The main specular spot is very intense compared to the two side spots. This is the typical signature of TiO$_2$ terminated STO \cite{Koster2000}. Kikuchi lines are visible as well, indicating the smoothness of the substrate. The side spots are doubled, which is due to the additional periodicity at the surface from the regularly spaced terrace steps. Figure~\ref{rheedini}, top right, shows the RHEED image of the LSMO film after the deposition of 5 unit cell layers. Clear two dimensional spots are visible, but also 2D streaks are present. The latter are attributed to the scattering of the RHEED beam off the unit cell high steps at the surface. Similar RHEED images were observed after the completion of films with thicknesses up to 40 nm. 

The main graph in figure~\ref{rheedini} shows the intensity of the specular reflection as measured during the initial growth of LSMO. The intensity shows oscillations which correspond to the growth of the individual unit cell layers. Within the oscillations, recovery of the intensity after the sudden decrease during the laser burst is observed, as shown more clearly in the inset of the graph. The oscillation amplitude decreases with the amount of material deposited during the first part of the growth. This decrease of the intensity is due to the difference in reflectivity of the STO surface and the LSMO surface, due to the increased scattering from the heavy La ions in the lattice. Finally, the full width at the half maximum (FWHM) of the intensity of the specular spot, measured along the (10) direction, is presented as well. During the growth, the RHEED spots are periodically more streaky, because more stepedge scattering is present at half unit cell layer coverage compared to full unit cell layer coverage. Therefore, the FWHM oscillates as well. The FWHM depends only on the shape of the intensity distribution and not on the total intensity and it is therefore a better indicator of the surface morphology than the main intensity of the reflection. As can be seen in the figure, the FWHM during the growth is almost equal to the FWHM of the substrate reflection indicating a smooth surface morphology. From these measurements it is concluded that the initial stage of the LSMO growth proceeds in the ideal 2D layer-by-layer growth mode.   

During the growth of the film, an increase of the RHEED oscillation maximum intensity was observed, as presented in figure~\ref{rheed25}. The oscillation intensity maximum typically peaked around 20 to 25 unit cell layers. The origin of the maximum in intensity is not understood. The maximum in intensity was not observed during growth on an $A$-site terminated substrate surface (SrRuO$_3$ film grown on STO \cite{Rijnders2004}), so it might indicate a termination conversion. After this peak the oscillation intensity maximum decreased rapidly and the oscillation amplitude became comparable to the intensity variations in the laser pulse recovery cycles. It is concluded that the LSMO growth mode during the latter part of the growth is close to the steady state growth mode which is characterized by a relatively constant step density, which is large compared to the step density of the initial substrate surface. 

\subsection{Surface morphology}
\begin{figure}
\centering
\includegraphics*[width=10cm]{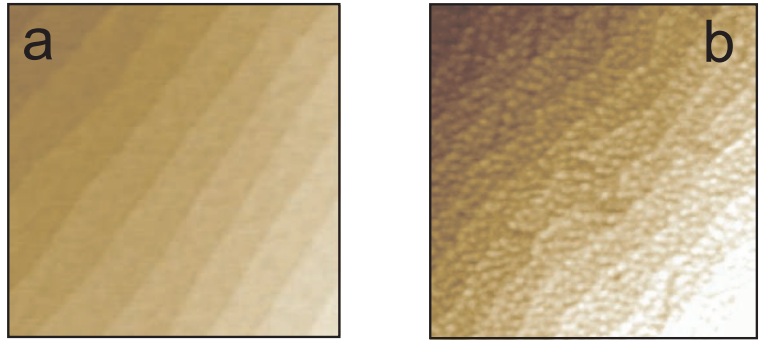}
\caption{Surface morphology as measured with \textit{ex situ} AFM. a) 20 unit cell layers of LSMO grown on STO. b) 30 unit cell layers of LSMO grown on STO. Both images are 2 by 2 $\mu$m.}
\label{afm}
\end{figure}

The surface morphology of the films was studied with atomic force microscopy (AFM). The images showed smooth surfaces as shown in the examples presented in figure~\ref{afm}. The step and terrace structure of the substrate is still observed on the surface of the films. Figure~\ref{afm}a shows the surface morphology of a 20 unit cell layer film which was completely grown in the 2D layer-by-layer growth mode, while figure~\ref{afm}b shows the surface morphology of a 30 unit cell layer film where the growth changed to the steady state growth mode. For the latter film islands with unit cell height are observed on the terraces. The root mean square roughness of the samples is 0.15 nm (20 uc sample) and 0.21 nm (30 uc sample) respectively. For LSMO, it is known that off-stoichiometric deposition results in precipitate formation on the surface \cite{Higuchi2009}. Such precipitates have not been observed.
 
\section{Crystal structure}
\label{thinfilmcrystalstructure}
\begin{figure}
\centering
\includegraphics*[width=8cm]{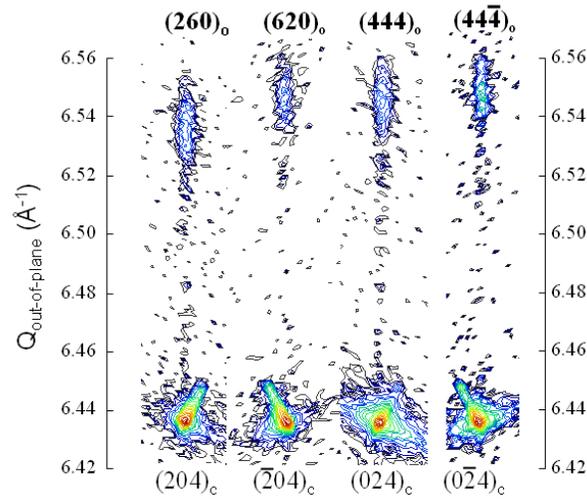}
\caption{Room temperature XRD RLMs around the (260)$_\textrm{o}$, (444)$_\textrm{o}$, (620)$_\textrm{o}$ and (44$\overline{4}$)$_\textrm{o}$ reflections of a 40 nm LSMO film grown on STO.}
\label{xrdrlm}
\end{figure}

\begin{figure}
\centering
\includegraphics*[width=8cm]{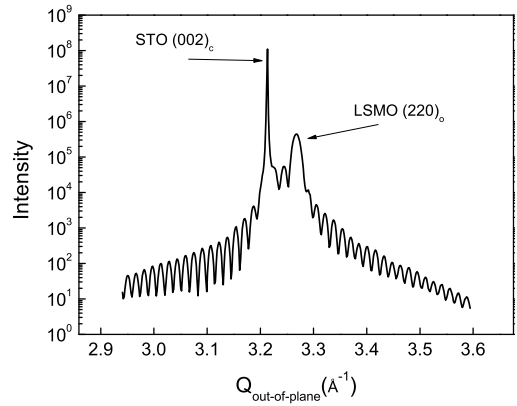}
\caption{Room temperature XRD HK scan (along the out-of-plane direction) of a 40 nm LSMO/STO sample. The substrate and film Bragg peaks have been indicated and multiple thickness fringes are observed.  }
\label{xrdt2t}
\end{figure}

The crystal structure of the LSMO thin films was analyzed with the use of x-ray diffraction (XRD) measurements. The measurements were performed using a PANalytical X'Pert materials diffractometer or at the beamline 7-2 of the Stanford Synchrotron Radiation Laboratory. 

The LSMO bulk rhombohedral crystal structure \cite{Radaelli1997} is incommensurate with the square/rectangular surface unit cells of the substrates. The LSMO is therefore under both normal strain as well as shear strain \cite{Farag2005}. To accommodate this strain LSMO adopts a different crystal symmetry, namely a distorted (monoclinic) orthorhombic unit cell with space group P2$_1$/m \cite{VailionisPRB}. The orthorhombic unit cell is (110)$_\textrm{o}$ oriented out-of-plane with [1$\overline{1}$0]$_\textrm{o}$ and [001]$_\textrm{o}$ in-plane orientations. To confirm that the LSMO layers were grown in a fully coherent fashion with respect to the underlying substrate, reciprocal lattice maps (RLM) were taken around symmetrical and asymmetrical reflections. 

As an example, RLMs around the (260)$_\textrm{o}$, (444)$_\textrm{o}$, (620)$_\textrm{o}$ and (44$\overline{4}$)$_\textrm{o}$ reflections of a 40 nm LSMO film grown on STO are shown in figure~\ref{xrdrlm}. The equal in-plane momentum transfer of the film and substrate peaks indicates a fully coherent film. The very small difference in the (260)$_\textrm{o}$ and (620)$_\textrm{o}$ atomic plane spacings represents a small difference in the a$_\textrm{o}$ and b$_\textrm{o}$ film lattice parameters. The refined lattice parameters are as follows: $\mathbf{a}$$_\textrm{o}$=5.480~\AA, $\mathbf{b}$$_\textrm{o}$=5.483~\AA, $\mathbf{c}$$_\textrm{o}$=7.809~\AA, $\alpha_\textrm{o}$=$\beta_\textrm{o}$=90$^\circ$ and $\gamma_\textrm{o}$=90.87$^\circ$. The pseudocubic lattice parameters are $\mathbf{a}$$_\textrm{pc}$=$\mathbf{b}$$_\textrm{pc}$=3.905~\AA~and $\mathbf{c}$$_\textrm{pc}$=3.846~\AA. The errors in the lattice parameters are 0.001~\AA~and 0.01$^\circ$. 

Figure~\ref{xrdt2t} shows an HK scan of the (220)$_\textrm{o}$ LSMO and (001)$_\textrm{c}$ STO Bragg peaks. Next to the peak, finite thickness fringes are observed whose period corresponds well to the 40 nm film thickness as estimated from counting the RHEED oscillations. The clear oscillations indicate a high quality sample with a smooth substrate film interface and a smooth film surface, as was also concluded from the AFM measurements. 

\section{Functional properties}
\label{chap3func}
The functional properties of LSMO are its magnetization and electrical transport. Here, measurements are presented of an 11.5 nm LSMO film grown on STO (001)$_\textrm{c}$. The sample thickness was determined by counting RHEED oscillations and was confirmed with an XRD reflectivity measurement. The sample has a Curie temperature of 350~K, a low temperature saturation magnetization of 4.0$\pm$0.05 $\mu_\textrm{B}$/Mn and a low temperature resistivity of 70 $\mu\Omega$cm.

\subsection{Magnetization}

\begin{figure}
\centering
\includegraphics*[width=10cm]{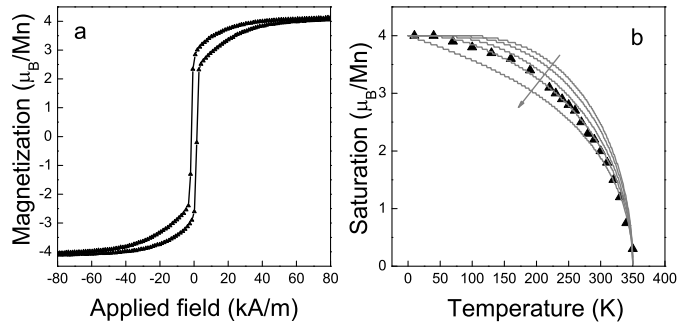}
\caption{Magnetization measurements of 11.5 nm LSMO grown on STO (001)$_\textrm{c}$. a) Hysteresis loop obtained at 10 K with the field applied in the plane of the film parallel to the [100]$_\textrm{pc}$ LSMO lattice direction. b) Temperature dependence of the saturation magnetization. Datapoints are indicated with black triangles and the curves are calculated from the Brillouin functional dependence of a Weiss ferromagnet. The shape of the curve depends on the value of $J$, $J$=$\frac{1}{2}$,1,2,4 and $\infty$ respectively, following the direction of the arrow. }
\label{magn}
\end{figure}

The magnetization of the sample was measured with a vibrating sample magnetometer (VSM) (Physical Properties Measurement System (PPMS) by Quantum Design). Figure~\ref{magn}a shows the low temperature magnetization loop, obtained at 10 K with the field applied along the [100]$_\textrm{pc}$ axis and after subtraction of a diamagnetic background from the substrate. The saturation magnetization is reached for an applied field strength larger than 50 kA/m ($\approx$630 Oe) and is equal to 4.0$\pm$0.05 $\mu_\textrm{B}$/Mn. This implies that next to the expected 3.7 $\mu_\textrm{B}$/Mn spin angular momentum also 0.3 $\mu_\textrm{B}$/Mn orbital angular momentum is present in the LSMO. Saturation values of 4 or more $\mu_\textrm{B}$/Mn have been observed earlier in single crystals with doping $x$=0.1 and $x$=0.2 \cite{Urushibara1995} and in LSMO polycrystals \cite{Koide2001}. The magnetization at remanence, 2.7 $\mu_\textrm{B}$/Mn, is significantly smaller than the saturation magnetization. This is partially because the measurement is along the magnetic hard axis \cite{Mathews2005}, and partially because the weak magnetic anisotropy of LSMO on STO results in nanoscale domain formation \cite{Houwman2008}. The magnetization is very soft with a coercivity of 1.6 kA/m ($\approx$20 Oe). 

The temperature dependence of the magnetization is presented in figure~\ref{magn}b. For each datapoint in the graph, a full hysteresis loop between 240 and -240 kA/m ($\approx$3000 Oe) was measured and the saturation magnetization was calculated after the background subtraction. The Curie temperature of the sample is 350 K. The saturation magnetization follows the Brillouin functional dependence for a Weiss ferromagnet \cite{Chikazumi1964}. The curve calculated using $J$=4, $J$ is the total angular momentum, results in the best fit of the data. This indicates either that the spins in the LSMO behave not completely classically (the case of $J$=$\infty$) and not completely quantum mechanically ($J$=$\frac{1}{2}$), but show intermediate behaviour or that spin wave excitations play an important role in the temperature dependence of the magnetization.

\subsection{Electrical transport}

\begin{figure}
\centering
\includegraphics*[width=8cm]{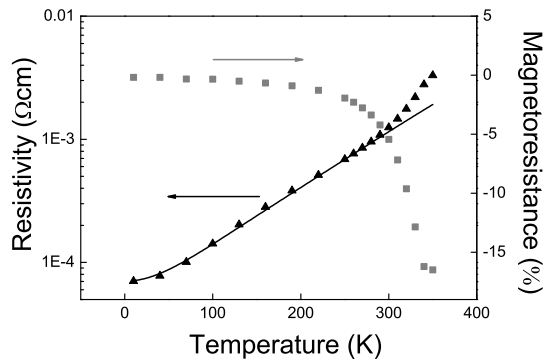}
\caption{Electrical transport measurements of 11.5 nm LSMO grown on STO (001)$_\textrm{c}$. Temperature dependent resistivity and magnetoresistance. The solid black line is a fit to the low temperature resistivity data, as discussed in the main text.}
\label{res}
\end{figure}

The resistivity of the samples was measured in the van der Pauw configuration \cite{vdPauw1958} (PPMS by Quantum Design). In order to obtain ohmic contacts between the aluminium bonding wires and the LSMO layer, gold contacts were deposited on the corners of the sample with the use of a shadow mask. Measurements were performed as a function of temperature (10-350~K) and magnetic field (0-2.5 T). The temperature dependent resistivity of the sample is shown in figure~\ref{res}. The sample shows metallic behaviour with a residual resistivity at 10 K of 70 $\mu\Omega$cm. For thicker films ($\ge$20 nm), a residual resistivity of 60 $\mu\Omega$cm was measured. The low temperature part of the resistivity curve can be described with the relation:
\begin{equation}
\rho(T) = \rho_0 + a T^2 + b T^5,
\label{resist}
\end{equation}
in which $\rho_0$ is the temperature independent impurity scattering and the $T^2$ ($T^5$) term describes electron electron (electron phonon) scattering. A fit using equation~\ref{resist} is shown in the figure and it decribes the curve up to a temperature of 300 K well. At 300 K a discontinuity in the slope of the resistivity curve is observed and above 300 K the resistivity increases faster compared to the model, indicating the approach of the metal insulator transition at $T_\textrm{C}$.

Figure~\ref{res} also presents the magnetoresistivity as a function of temperature. At each temperature the magnetic field is swept between -2.5 and 2.5~T and the magnetoresistivity at 2.5~T $(\rho(2.5T)-\rho(0))/\rho(0)$ is shown. The magnetoresistance is negative and largest, -16 \%, at the Curie temperature. Hall effect measurements determined the charge carriers to be holes, as expected for LSMO. The observed Hall coefficient $R_\textrm{H}$=4.6$\cdot10^{-4}$ cm$^3$/C implies a carrier density of approximately 0.8 hole/unit cell, which is in reasonable agreement with earlier measurements on single crystals \cite{Asamitsu1998}.   

\section{Photoelectron spectroscopy}
Several LSMO thin films were characterized with \textit{in situ} x-ray and ultra violet photoelectron spectroscopy (XPS and UPS). The measurements were performed with an XPS/UPS system designed by Omicron Nanotechnology GmbH, equipped with an EA 125 electron energy analyzer. For XPS an Al K$\alpha$ source (XM 1000) was used and the UV light source is a He plasma lamp (HIS 13) operated at the HeI (21.22 eV) excitation edge. The base pressure of the system was below 10$^{-10}$ mbar. The analyzer was calibrated with the use of an \textit{in situ} sputter cleaned Au sample. 

\begin{figure}
\centering
\includegraphics*[width=8cm]{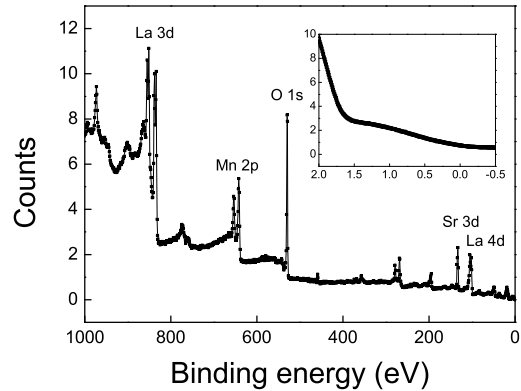}
\caption{An XPS survey scan of a 10 unit cell layer LSMO sample grown on (001)$_\textrm{c}$ STO. The La 3d, Mn 2p, O 1s, Sr 3d and La 4d peaks are indicated. The inset shows the photoelectron intensity in the region close to the Fermi energy measured with UPS. The measurements were performed at room temperature.}
\label{xpsoverview}
\end{figure}

Figure~\ref{xpsoverview} shows a survey scan of a 10 unit cell layer LSMO sample grown on (001)$_\textrm{c}$ STO. The main features include the La 3d peaks at 850 $e$V, the oxygen 1s peak at 531 $e$V, the Mn 2p peaks at 640 $e$V and the Sr and La peaks at low binding energy. Next to this a small Ti peak originating from the substrate can be seen at 460 $e$V. No indications for the presence of impurity atoms, including carbon contamination, was found. This is attributed to the \textit{in situ} measurement. No attempts have been made to quantify the composition of the film, as large uncertainty margins exist for the element sensitivity factors required to normalize measured peak intensities. From XRD and functional properties measurements it is clear that the composition of the films is equal to the desired LSMO composition within the rather large error margin of the XPS measurements.

The inset of figure~\ref{xpsoverview} shows the valence band spectrum as measured with UPS at room temperature. No Fermi edge is observed, the photoelectron count increases linearly with binding energy up to a value of 1.5 $e$V, after which a steep increase is observed. The absence of the Fermi edge is not surprising. In literature, several studies have observed a Fermi edge, but only at low ($<$50K) temperature \cite{Parknat1998, Dessaucmobook} or using resonant photoemission spectroscopy \cite{Horiba2005}. LSMO's electron density at the Fermi energy is much reduced compared to band theory due to the presence of the pseudogap \cite{Dessaucmobook, Chuang2001} and disappears above the Curie temperature. It is therefore natural that the Fermi edge cannot easily be observed in a room temperature measurement. The steep increase of the electron density below 1.5 $e$V is in good agreement with the literature \cite{Parknat1998}.

Several authors have suggested that Sr or Ca segregation towards the surface of the film is an intrinsic effect in LSMO (or La$_{0.67}$Ca$_{0.33}$MnO$_3$ (LCMO) respectively) thin films \cite{Simon2004, Herger2008}. Angle dependent XPS measurements can be used to study changes in the composition of a thin film in the surface layer. The results obtained for a 10 unit cell layer film grown on (001)$_\textrm{c}$ STO are presented in table~\ref{HBtable}. A film thickness of ten unit cell layers was chosen for this experiment in order to directly compare with the 1 to 9 unit cell layer experiments by Herger  \textit{et al.}\cite{Herger2008}. If thinner LSMO layers would be used the XPS Sr signal would gain relative intensity due to a substrate contribution, affecting the measurement. As the relative intensity of peaks measured at different take-off angles does not depend on the element sensitivity factors, quantitative analysis is possible. The Sr 3d peak at 135 $e$V was compared with the La 4d peak at 105 $e$V and for both elements, an intensity ratio between bulk and surface sensitive measurements of 1.5$\pm$0.05 was found. This limits the off-stoichiometry of the surface to within 6\%. In the example in literature \cite{Herger2008} it is mentioned that the topmost layer has a La/Sr ratio of 0.5 instead of 2, while the subsequent layers are stoichiometric. Based on an electron escape depth of 2 nm for electrons with kinetic energy of 1350 $e$V \cite{nist}, Herger's 9 unit cell layer sample would show an off-stoichiometry of 33\% in the XPS measurement. Therefore, it is concluded that Sr segregation is much reduced in these samples as compared to the samples of Herger \textit{et al.}. It is known that the Sr segregation depends on the oxygen partial pressure \cite{Fister2008}. Therefore, the lack of Sr segregation can be attributed to the high oxygen pressure used during the deposition.

\begin{table}
\centering
\caption{Integrated XPS peak intensity for a 10 unit cell layer LSMO sample grown on (001)$_\textrm{c}$ STO. The XPS intensity is in arbitrary units and for peaks split by e.g. spin orbit coupling the intensities of all the components have been added. The error margins are estimated on the quality of the background subtraction routines. }
\begin{tabular}{ccccc}
\hline
\\[-1.5ex]
Peak & Binding energy ($e$V) &10$^\circ$	Intensity & 60$^\circ$ Intensity 	&	Ratio		\\
 & & (bulk sensitive) & (surface sensitive) &\\
\\[-1.5ex]
\hline
\\[-1.5ex]
Sr 3d	& 135 & 48	$\pm$0.5								& 32			$\pm$0.5					&	1.5 $\pm$ 0.05						\\
La 4d	& 105 & 90$\pm$1										& 60		$\pm$1						&	1.5 $\pm$ 0.05						\\
\\[-1.5ex]
\hline
\end{tabular}
\label{HBtable}
\end{table}

\section{Scanning transmission electron microscopy}

\begin{figure}
\centering
\includegraphics*[width=10cm]{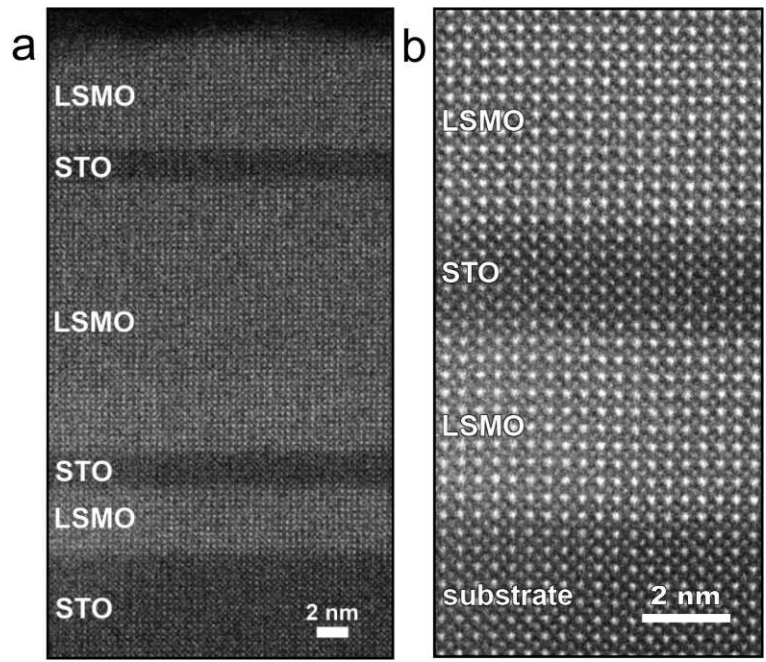}
\caption{HAADF-STEM images of an LSMO/STO heterostructure. a) Low magnification overview. b) High magnification image of the area close to the substrate. }
\label{stem}
\end{figure}

A multilayer LSMO/STO sample was characterized with scanning transmission electron microscopy (STEM), to study the interface atomic structure. The STEM data presented was measured with a FEI Titan microscope. The sample is comprised of several layers, from substrate to surface: 4 nm LSMO, 2 nm STO, 16 nm LSMO, 2 nm STO and 8 nm LSMO. 

Figure~\ref{stem}a presents a low magnification HAADF image of the different layers in the sample. In HAADF microscopy, the observed intensity of a column of atoms scales with the atomic weight of the elements within the columns. Therefore, the LSMO is brighter in the image than the STO. Figure~\ref{stem}b presents a higher magnification image obtained from the layers close to the substrate. Both the $A$- and $B$-site columns are observed. The multilayer structure is grown coherently and following a quantitative analysis by statistical parameter estimation \cite{Aert2009}, it can be concluded that the interfaces are well defined with a chemical roughness of maximum 1 unit cell.

\section{Comparison to other groups}
\label{lsmooverview}

\begin{table}
\centering
\caption{Overview of the thin film properties. $T_\textrm{C}$ is the Curie temperature, $\rho_0$ is the residual resistivity at 10 K and $M_\textrm{sat}$ is the saturation magnetization at 10 K. }
\begin{tabular}{cccc}
\hline\hline
\\[-1.5ex]
Group& [001]$_\textrm{pc}$ axis (\AA) & $T_\textrm{C}$ (K) & \\
\\[-1.5ex]
Deposition 	&	$\rho_0$ ($\mu\Omega$cm) &$M_\textrm{sat}$ ($\mu_\textrm{B}$/Mn) & 	References	\\
\\[-1.5ex]
\hline\hline

\\[-1.5ex]
Tokyo (Tokura \textit{et al.})& - & 370 & 								\\
\\[-1.5ex]
Single crystal	& 90								& 3.5				&	\cite{Urushibara1995}						\\
\\[-1.5ex]

\hline
\\[-1.5ex]
Tsukuba (Kawasaki \textit{et al.})& 3.83 & 340 & 							\\
\\[-1.5ex]
 PLD	& 200								& 3.6			&	\cite{Izumi1998, Yamada2004, Horiba2005, Yamada2006}						\\
\\[-1.5ex]
\hline

\\[-1.5ex]
Illinois (Eckstein \textit{et al.})& 3.85 & 355 & 									\\
\\[-1.5ex]
 MBE	& 40								& 3.2				&	\cite{Odonnell2000, Kavich2007, Bhattacharya2008}						\\
\\[-1.5ex]

\hline

\\[-1.5ex]
Orsay Cedex (Contour \textit{et al.})	& - & 350 &									\\
\\[-1.5ex]
 PLD	& -								& 3.7				&	\cite{Maurice2002, Pailloux2002, Bowen2003}						\\
\\[-1.5ex]
\hline

\\[-1.5ex]
Rome (Balestrino \textit{et al.})&3.85 & 300  & 								\\
\\[-1.5ex]
 PLD	& 1000								& 3.5				&	\cite{Tebano2006, Aruta2007,Tebano2008}						\\
\\[-1.5ex]
\hline
\\[-1.5ex]
Tokyo (Hwang \textit{et al.})	& 3.84& 360 & 									\\
\\[-1.5ex]
 PLD	& 60								& -				&	\cite{Song2008, Kim2010, Fitting2010}						\\
\\[-1.5ex]
\hline

\\[-1.5ex]
Twente (This work)	& 3.845	& 350 & 									\\
\\[-1.5ex]
 PLD	& 60								& 4.0				&	\cite{Huijben2008, Gray2010}						\\
\\[-1.5ex]

\hline\hline
\end{tabular}
\label{bigtable}
\end{table}

Table~\ref{bigtable} presents an overview of the most important properties of LSMO thin films. Data taken from different research groups is included in order to compare the thin film quality. For reference also the properties of a single crystal are mentioned. All films were grown on STO (001)$_\textrm{c}$ substrates by PLD, except for the samples grown in Illinois, for which MBE was used. Most groups published AFM measurements of the surface morphology, which showed smooth surfaces with the step and terrace structure of the substrate clearly visible. Two groups, the Illinois group and the Rome group, did not publish images of the film surface morphology, but they did present a RHEED image from which it was concluded that the films were smooth \cite{Kavich2007, Tebano2006}. The Rome group mentions that their films have a metal-insulator transition temperature of 370 K. This temperature is defined in their papers as the temperature at which the resistivity of the films has a maximum and it is not equal to the Curie temperature. The value mentioned in the table is deduced from the temperature dependence of the magnetization plot in \cite{Tebano2008}. The data from the 20 unit cell layer sample was used, which has a similar metal-insulator transition temperature as their thick samples.

In general, a similar crystal structure with an out-of-plane lattice parameter of 3.83 to 3.85~\AA~and a similar surface morphology is obtained. Nevertheless, differences in the functional properties are present. We therefore suspect that these differences are due to small variations in the stoichiometry of the films, e.g. the La/Sr ratio and the degree of oxygenation of the film.

A high quality sample should combine a high $T_\textrm{C}$ with a high saturation magnetization and a low residual resistivity. The samples described here have the highest saturation magnetization, a high $T_\textrm{C}$ and a low residual resistivity. Only the samples from the Illinois group have a lower residual resistivity while both the Illinois and the Tokyo group obtained a slightly higher $T_\textrm{C}$. We obtained a higher saturation magnetization, 4.0 $\mu_\textrm{B}$/Mn in contrast to the 3.2 $\mu_\textrm{B}$/Mn reported by the Illinois group. The Tokyo group did not report the saturation magnetization, but the magnetization value at 200 K can be compared; 2.7 $\mu_\textrm{B}$/Mn (Tokyo \cite{Song2008}) versus 3.2 $\mu_\textrm{B}$/Mn (this work). Combining all three properties it is concluded that the samples of this work combine both a high magnetization with a low residual resistivity and have the highest quality reported so far in literature. 

\section{Conclusion}
In this article, an overview of the fabrication and the properties of high quality LSMO thin films was presented. The films have a saturation magnetization of 4.0 $\mu_\textrm{B}$/Mn, a Curie temperature of 350 K and a residual resistivity of 60 $\mu\Omega$cm. XPS measurements show the absence of chemical impurities and no evidence for surface segregation. STEM measurements show that multi-layer structures with sharp interfaces were realized. These results indicate that high quality films, combining both large magnetization and small residual resistivity, were realized. For LSMO devices, the thin film quality is crucial for the device performance. Therefore, this research is important for the application of LSMO in devices.  

\ack
This research was financially supported by the Dutch Science Foundation, by NanoNed, a nanotechnology program of the Dutch Ministry of Economic Affairs and by the NanOxide program of the European Science Foundation. This work is supported in part by the Department of Energy, Office of Basic Energy Sciences, Division of Materials Sciences and Engineering, under contract DE-AC02-76SF00515.

\section*{References}
\bibliographystyle{unsrt}

\end{document}